\documentclass[english,aps,reprint]{revtex4-2}
\usepackage[T1]{fontenc}
\usepackage[latin9]{inputenc}
\usepackage{bm}
\usepackage{amsmath}
\usepackage{amsthm}
\usepackage{amssymb}

\makeatletter
\usepackage{newtxtext}

\usepackage{babel}

\allowdisplaybreaks
\usepackage{bbm}
\usepackage{stmaryrd}

\newcommand*{\boldone}{\text{\usefont{U}{bbold}{m}{n}1}}

\makeatother

\usepackage{babel}
\begin{document}
\title{Thermodynamic dissipation does not bound replicator growth and decay
rates}
\author{Artemy Kolchinsky}
\email{artemyk@gmail.com}

\affiliation{ICREA-Complex Systems Lab, Universitat Pompeu Fabra, 08003 Barcelona,
Spain}
\affiliation{Universal Biology Institute, The University of Tokyo, 7-3-1 Hongo,
Bunkyo-ku, Tokyo 113-0033, Japan\looseness=-1}
\begin{abstract}
In a well-known paper, Jeremy England derived a bound on the free
energy dissipated by a self-replicating system {[}England, ``Statistical
physics of self-replication'', \emph{The Journal of Chemical Physics},
2013{]}. This bound is usually interpreted as a universal relationship
that connects thermodynamic dissipation to replicator per-capita decay
and growth rates. We argue from basic thermodynamic principles against
this interpretation. In fact, we suggest that such a relationship
cannot exist in principle, because it is impossible for a thermodynamically-consistent
replicator to undergo both per-capita growth and per-capita decay
back into reactants. Instead, replicator may decay into separate waste products, 
but in that case, replication and
decay are two independent physical processes, and there is no universal
relationship that connects their thermodynamic and dynamical properties.
\end{abstract}
\maketitle
\global\long\def\tto{\shortrightarrow}%
\global\long\def\SS{A}%
\global\long\def\ss{a}%
\global\long\def\mO{\bm{\mathrm{0}}}%
\global\long\def\mII{\bm{\mathrm{II}}}%
\global\long\def\mI{\bm{\mathrm{I}}}%
\global\long\def\mII{\bm{\mathrm{II}}}%
\global\long\def\perhr{/\text{hour}}%
\global\long\def\dmin{\delta_{\min}}%
\global\long\def\ddeath{\delta_{\text{degrade}}}%
\global\long\def\dhydro{\delta_{\text{hydro}}}%
\global\long\def\duc{\delta_{\text{uncopy}}}%
\global\long\def\dproton{\delta_{\text{proton}}}%
\global\long\def\stot{\Delta s_{\text{tot}}(\mI\tto\mII)}%
\global\long\def\stotBase{\Delta s_{\text{tot}}}%
\global\long\def\sint{\Delta s_{\text{int}}}%
\global\long\def\npep{n_{\text{pep}}}%
\global\long\def\stotMO{\Delta s_{\text{tot}}(\mI\tto\mO)}%
\global\long\def\kk{\kappa}%
\global\long\def\kB{k_{B}}%
\global\long\def\dtt{\tau}%
\global\long\def\replicatorEP{\sigma}%
\global\long\def\microstate{x}%
\nocite{england2013statistical}

\section{Introduction}

Research in thermodynamics has shown that there are universal relationships
between the thermodynamic and dynamic properties of nonequilibrium
processes. The most famous relationship, termed \emph{local detailed
balance} (LDB), says that the statistical irreversibility of a microscopic
process is directly related to the thermodynamic dissipation, i.e.,
the entropy production in the system and the environment during that
process \citep{maes2021local}. The generality of LDB hints at the
possibility of universal bounds on the thermodynamic properties of
living systems.

This idea inspired a 2013 paper by England on the thermodynamics
of self-replicating systems~\citep{england2013statistical}. Consider
a population of replicators with per-capita replication rate $g$
and per-capita decay rate $\delta$, where decay is defined as the
``reversion of the replicator back into the exact set of reactants
in its environment out of which it was made'' \citep{england2013statistical}.
A system with fixed per-capita rates of replication $g$ and decay
$\delta$ may be said to exhibit \emph{first-order growth} and \emph{first-order
decay}. {The stochastic dynamics of first-order growth and
decay may be described by a Markovian birth-death master equation,
\[
\dot{p}_{n}(t)=(n-1)gp_{n-1}(t)+(n+1)\delta p_{n+1}(t)-n(g+\delta)p_{n}(t),
\]
where $p_{n}(t)$ is the probability that the population contains
$n$ replicators at time $t$. For large population sizes $n$, this
master equation may be approximated as \citep{saakian2016nonlinear}}
\begin{equation}
\dot{p}_{n}(t)\approx ng(p_{n-1}(t)-p_{n}(t))-\delta n(p_{n}(t)-p_{n+1}(t)),\label{eq:me}
\end{equation}
which appears as Eq.~(9) in Ref.~\citep{england2013statistical}.
Finally, neglecting fluctuations for large $n$, the expected population
size at time $t$  will grow
exponentially as
\begin{equation}
\langle n\rangle_{p(t)}\approx\langle n\rangle_{p(0)}e^{(g-\delta)t}.\label{eq:exp}
\end{equation}
Eq.~\eqref{eq:exp} relates per-capita replication and decay rates
in the master equation to long-term population dynamics. 

The main result of England's paper~\citep{england2013statistical} [Eq.~(10)] is a thermodynamic bound on the
ratio of growth and decay rates,
\begin{equation}
\stotBase\ge\ln\frac{g}{\delta}\,,\label{eq:bnd1}
\end{equation}
where $\stotBase$ is the entropy production incurred when a single
replicator makes a copy of itself. The quantity $\stotBase$ is proportional
to the nonequilibrium free energy dissipated during replication \citep{parrondoThermodynamicsInformation2015}.
England illustrates the bound using two real-world systems of interest:
an RNA-based molecular replicator constructed by Lincoln and Joyce
\citep{lincolnSelfsustainedReplicationRNA2009} and an \emph{E. coli}
bacterium.

The bound (\ref{eq:bnd1}) appears to bridge two different worlds:
the physical world of thermodynamic dissipation and the biological
world of replicator dynamics.  From an intellectual perspective,
we find England's proposal stimulating and elegant. However, by studying
the thermodynamics of simple molecular replicators \citep{kolchinsky2024thermodynamic},
we have come to find that the bound~\eqref{eq:bnd1} must be interpreted
with great care.

In this paper, we argue that --- contrary to standard interpretations
of England's result --- inequality~\eqref{eq:bnd1} does not provide
a thermodynamic bound on the growth and decay rates of replicators. 
First, in Section~\ref{sec:Impossibility-theorem}, we derive a kind of ``impossibility
theorem''. It shows that no bound like~\eqref{eq:bnd1} can apply to self-replicating systems, because a replicator cannot undergo both first-order growth
and first-order decay back into reactants without violating the laws
of thermodynamics. Rather, as we discuss in Section~\ref{sec:alt},
two options present themselves. First, a replicator may decay via 
the reverse process of autocatalysis, in which case decay will not
be first-order. Alternatively, a replicator may undergo first-order
decay into a different set of waste products, in which case there is no universal relationship between the physical properties of the two independent processes of replication and decay. 

{We emphasize that our analysis employs the 
same setup and theoretical framework (stochastic thermodynamics) as does England's paper. Thus, our work is not meant as a critique of the general approach of stochastic thermodynamics, but rather of a particular application to replicating systems.}  

Before proceeding to our analysis, we first review the derivation
of England's result, in the process filling in a few subtle details.

\section{Background}

We begin by reviewing the setup and derivation of England's bound
\eqref{eq:bnd1} in Ref.~\citep{england2013statistical}. 
We proceed
in several steps. In Section~\ref{subsec:General-result}, we derive
a ``weak'' version of LDB that applies at the level of coarse-grained
macrostates. This is a general result that relates dissipation and
statistical irreversibility for arbitrary choices of macrostates and
systems, whether replicating or not. In Section~\ref{subsec:special-case},
we derive England's bound by applying the weak version of LDB to the
special case of self-replicating systems. In Section~\ref{subsec:gen},
we derive a generalized version of England's bound, which will be
useful for our analysis below.

\subsection{Local detailed balance and coarse-graining}

\label{subsec:General-result}

\global\long\def\PpYXs{P_{\tau}(y\tto x)}%
\global\long\def\PpXYs{P_{\tau}(x\tto y)}%
\global\long\def\PpYX{P_{\tau}(y\!\tto\!x)}%
\global\long\def\PpXY{P_{\tau}(x\!\tto\!y)}%
\global\long\def\microP{P}%
\global\long\def\pI{\microP_{\mI}}%
\global\long\def\pII{\microP_{\mII}}%
\global\long\def\QXY{Q(x\tto y)}%
\global\long\def\PItoII{\pi(\mI\!\tto\!\mII)}%
\global\long\def\PIItoI{\pi(\mII\!\tto\!\mI)}%

Consider a system coupled to a heat bath at temperature $T$ which
undergoes an undriven (time-independent) process over time interval
$[t,t+\tau]$. The system's microscopic dynamics are described
by the conditional probability $\PpXY$ that the system ends in microstate
$y$ at time $t+\tau$, given that it started in microstate $x$ at
time $t$. We generally focus on chemical or biological systems in
solution that are overdamped, meaning that their
momentum degrees-of-freedom equilibrate quickly and can be ignored
at the microscopic level of description. 
{Each microstate $x$ might specify the position of each particle in the system. In other cases, the microstates may represent coarse-grained ``mesostates'', e.g., they may specify only the positions of the solute particles, the counts of different particle types in a well-mixed system, etc. In all cases, we require that each microstate $x$ is internally in equilibrium. Another requirement is that the microscopic dynamics are Markovian over timescale $\tau$, meaning that the probability of microstate at time $t+\tau$ depends only on the microstate at time $t$.}

The system possesses an equilibrium Boltzmann distribution,
\begin{align}
\microP_{\mathrm{eq}}(x)=\frac{1}{Z}e^{-E(x)/k_{B}T},
\end{align}
where $E(x)$ is the energy of microstate $x$. 
According to the principle
of ``detailed balance'', the forward and backward probability fluxes
across each microscopic transition $x\to y$ balance in equilibrium,
\begin{equation}
\microP_{\mathrm{eq}}(x)\PpXY=\microP_{\mathrm{eq}}(y)\PpYX.\label{eq:db}
\end{equation}
The ratio of the forward and backward transition probabilities can
be written as
\begin{equation}
\frac{\PpXY}{\PpYX}=\frac{\microP_{\mathrm{eq}}(y)}{\microP_{\mathrm{eq}}(x)}=e^{[E(x)-E(y)]/k_{B}T}.\label{eq:ldb00-1}
\end{equation}
By the First Law of Thermodynamics, the energy lost by the system
during the transition $x\to y$ is equal to the heat transferred to
the bath, $\QXY=E(x)-E(y)$, therefore
\begin{equation}
\frac{\PpXY}{\PpYX}=e^{\QXY/k_{B}T}.\label{eq:ldb00}
\end{equation}

Observe that $\QXY/k_{B}T$ is the increase in the thermodynamic entropy
of the heat bath during the transition $x\to y$. Eq.~\eqref{eq:ldb00}
is a special case of the general principle of ``local detailed balance''
(LDB), which says that the statistical irreversibility of microscopic
transitions is related to the increase in the thermodynamic entropy of
the environment.

Suppose that the system is associated with two macrostates $\mI$
and $\mII$, i.e., two subsets of microstates, which in principle
may be chosen arbitrarily. Macrostate $\mI$ is described by a probability
distribution $\pI(\microstate)$ over microstates with support
restricted to $\mI$. Macrostate $\mII$ is also described by a
probability distribution over microstates $\pII(y)$. This distribution
is defined by propagating the distribution $\pI$ under the microscopic
dynamics and then conditioning on membership in macrostate $\mII$:
\[
\pII(y)=\frac{\boldone_{\mII}(y)\int_{\mI}\pI(\microstate)\PpXY\,dx}{\int_{\mII}\int_{\mI}\pI(\microstate)P_{\tau}(x\to y^{\prime})\,dx\,dy^{\prime}}
\]
where $\boldone$ is the indicator function. {In principle, the distributions $\mI$ and $\mII$ may be arbitrarily far from internal equilibrium.}

Entropy production refers to the increase in the entropy of the system and its environment during a process. {In stochastic thermodynamics, the entropy production incurred when going from macrostate $\mI$ to macrostate $\mII$ is defined as~\cite{esposito2011second,parrondoThermodynamicsInformation2015}}
\begin{equation}
\stot=[S(\pII)-S(\pI)]+\langle Q\rangle_{\mI\to\mII}/\kB T,\label{eq:stot2}
\end{equation}
where $S(\pII)-S(\pI)$ is the increase in the system's Shannon entropy
and $\langle Q\rangle_{\mI\to\mII}$ is the average heat generated. 
Here we use the notation
\begin{equation}
\langle\cdots\rangle_{\mI\to\mII}=\frac{\int_{\mII}\int_{\mI}\,\pI(x)\PpXY\,\cdots\,dx\,dy}{\int_{\mII}\int_{\mI}\,\pI(x)\PpXY\,dx\,dy}\label{eq:not}
\end{equation}
to indicate the expectation of a trajectory-level quantity conditioned
on initial macrostate $\mI$ and final macrostate $\mII$.

To derive a bound on the entropy production, consider the conditional
probability $\PItoII$ that the final microstate belongs to macrostate
$\mII$, given that the initial microstate is drawn from $\pI$. Consider
also the conditional probability $\pi(\mII\to\mI)$ that the final
microstate belongs to macrostate $\mI$, given that the initial microstate
is drawn from $\pII$. As shown in Ref.~\citep{england2013statistical},
$\stot$ is bounded by the logarithmic ratio of these two transition
probabilities, 
\begin{equation}
\stot\ge\ln\frac{\PItoII}{\PIItoI}.\label{eq:ldb2}
\end{equation}

{The bound~\eqref{eq:ldb2} can be derived from the microscopic principle of LDB~\eqref{eq:ldb00}. To do so, let us write the
macrostate transition probabilities as
\begin{align}
\PItoII & =\int_{\mI}\int_{\mII}\pI(x)\PpXY\,dx\,dy\label{eq:mItoIIdef}\\
\PIItoI & =\int_{\mII}\int_{\mI}\pII(y)\PpYX\,dy\,dx\label{eq:mIItoIdef}
\end{align}
The ratio of these transition probabilities can be written as
\[
\frac{\PIItoI}{\PItoII}=\frac{\int_{\mII}\int_{\mI}\pI(x)\PpXY e^{\ln\frac{\pII(y)}{\pI(x)}-\ln\frac{\PpXYs}{\PpYXs}}\,dy\,dx}{\int_{\mI}\int_{\mII}\pI(x)\PpXY\,dx\,dy}
\]
Using the expression of LDB~\eqref{eq:ldb00} and the notation~\eqref{eq:not}
gives
\begin{align*}
\frac{\PIItoI}{\PItoII} & =\langle e^{-[\ln\frac{\pI(x)}{\pII(y)}+Q(x\to y)/k_{B}T]}\rangle_{\mI\to\mII}\\
 & \ge e^{-\langle\ln\frac{\pI(x)}{\pII(y)}+Q(x\to y)/k_{B}T\rangle_{\mI\to\mII}}=e^{-\stot}.
\end{align*}
where the second line uses Jensen's inequality and then Eq.~\eqref{eq:stot2} 
(note that $\langle\ln\frac{\pI(x)}{\pII(y)}\rangle_{\mI\to\mII}=S(\pII)-S(\pI)$).
The bound~\eqref{eq:ldb2} follows by taking logarithms and rearranging.}

{The bound~\eqref{eq:ldb2} is a general inequality that
applies to various choices of macrostates and systems, whether replicating
or not. It may be interpreted as a ``weak'' version of LDB that
applies to transitions between coarse-grained macrostates. It
is weak in the sense that it relates statistical irreversibility and
thermodynamic dissipation via an inequality, in contrast to the stronger
version of LDB~\eqref{eq:ldb00} that holds as an equality at the
microscopic level.}

The weak version of LDB~\eqref{eq:ldb2} is derived using
Jensen's inequality, and it turns into equality when Jensen's inequality
is tight. In turn, Jensen's inequality is tight when the quantity $\ln\frac{\pI(x)}{\pII(y)}+Q(x\to y)/k_{B}T$
 (the fluctuating dissipation incurred by microscopic
trajectory $x\to y$) is the same for every trajectory that connects
macrostate $\mI$ to macrostate $\mII$. There are two ways this can
happen.

The first is that the macrostates are in internal equilibrium,
meaning that the macrostate distribution have a restricted Boltzmann
form,
\[
\pI(x)=\frac{1}{Z_{\mI}}\boldone_{\mI}(x)e^{-E(x)/k_{B}T}\qquad Z_{\mI}=\int_{\mI}e^{-E(x)/k_{B}T}\,dx,
\]
and similarly for $\pII$ and $Z_{\mII}$. With a bit of rearranging,
we find that the fluctuating dissipation is then given by
\[
\ln\frac{\pI(x)}{\pII(y)}+Q(x\to y)/k_{B}T=\ln Z_{\mII}-\ln Z_{\mI}.
\]
The quantity $\ln Z_{\mII}-\ln Z_{\mI}$ is proportional to the loss
of equilibrium free energy in going from macrostate $\mI$ to macrostate
$\mII$, and it does not depend on $x$ or $y$. In this case, there
are no fluctuations in the fluctuating dissipation, and the weak version of LDB
becomes an equality, $\stot=\ln\frac{\PItoII}{\PIItoI}=\ln Z_{\mII}-\ln Z_{\mI}$.

The second way that the weak version of LDB can be tight is when all
of the trajectories that connect macrostates $\mI$ and $\mII$ incur
the same dissipation, even though the two macrostates may not 
be in internal equilibrium. For instance, this may occur because there
is a single chemical reaction, or a single sequence of chemical reactions,
that transforms the system between macrostates $\mI$ and $\mII$.

Conversely, when there are large fluctuations in the dissipation incurred
by different trajectories $x\to y$, Jensen's inequality may be very
loose, therefore also the weak version of LDB may be very loose. Such fluctuations can
occur when the macrostates are not in internal equilibrium and there
exist multiple alternative pathways that go between macrostates $\mI$
and $\mII$. It is a well-known limitation of stochastic thermodynamics
that the exact relationship between statistical irreversibility and thermodynamic
dissipation only holds at the microscopic level \citep{maes2021local}.
It is also known that statistical irreversibility at the coarse-grained
level often underestimates the true amount of dissipation by many orders
of magnitude (for example, see Ref.~\citep{foster2023dissipation}).
We return to this issue in Section~\ref{sec:alt} below, when we
discuss thermodynamics bounds for replicating systems with multiple
degradation pathways.

\subsection{Application to replicating systems}

\label{subsec:special-case}

Following England \citep{england2013statistical}, we now apply the
weak version of LDB~\eqref{eq:ldb2} to the special case of replicating
systems. To do so, we define macrostate $\mI$ as the set of microstates
that contain a single replicator, along with any reactants needed
for successful replication. Macrostate $\mII$ is defined as the set
of microstates that contain two replicators: the parent replicator
found in macrostate $\mI$ and its new offspring, as well as any side
products that result from replication. Importantly, although the overall
system is not driven, macrostate $\mI$ may contain highly energetic
reactants that drive replication forward. 

The transition probability $\pi(\mI\to\mII)$ over time $\dtt$ is
approximated using the per-capita replication rate as $\pi(\mI\to\mII)\approx g\,\dtt$.
The transition probability $\pi(\mII\to\mI)$, corresponding to the
reversion of the new offspring back into ``the exact set of reactants
in its environment out of which it was made'', is approximated using
the per-capita decay rate as $\pi(\mII\to\mI)\approx\delta\,\dtt$.
This is not $2\delta\,\dtt$ because, in England's analysis, the parent and
offspring replicators are distinguished under $\mII$, and $\pi(\mII\to\mI)$
refers only to the decay of the new offspring (see also discussion
below~\eqref{eq:bnd1b}). Plugging these two approximations into~\eqref{eq:ldb2} and simplifying recovers England bound~\eqref{eq:bnd1}.

We note that England's bound involves the same per-capita growth $g$  and decay rates $\delta$ that appear in the birth-death master equation
\eqref{eq:me} that describes population-level dynamics. This birth-death equation specifies a continuous-time Markov chain, in which the  
the statistics of future replication and
decay events depend only on the current number of replicators, not on prior history. 
The assumption of Markovian population dynamics is standard
in population biology \citep{novozhilov2006biological} and chemistry
\citep{delbruck1940statistical,mcquarrie1963kinetics}, and it is
a reasonable starting point for relating dissipation 
to population dynamics. In principle, it may be justified under
a separation-of-timescales, where the timescales of division and internal
relaxation are shorter than waiting times between replication events.

\subsection{Generalization of England's bound}

\label{subsec:gen}

\global\long\def\sRep{\sigma_{\mathrm{rep}}}%

Observe that the term $\stotBase\equiv\stot$ in England's
bound~\eqref{eq:bnd1} refers specifically to the entropy produced when the system
goes from a macrostate with one replicator to a macrostate with two replicators.
Without additional assumptions, this does not necessarily equal the
entropy produced when the system goes from two to three replicators, three to
four replicators, etc. However, to study the thermodynamics of self-replicating
systems, we need a general expression for the entropy production
that holds at all population sizes. 

Here we introduce this general
expression for the entropy production, and we then use it to derive
a generalized form of England's bound. 
Specifically, we consider the entropy production when a system is
transformed from a macrostate with $n$ replicators to a macrostate
with $n^{\prime}$ replicators. As we show in the Appendix, this entropy
production is given by
\begin{equation}
\stotBase(n\to n^{\prime})\approx(n^{\prime}-n)\sRep+\ln\frac{n!}{n^{\prime}!}.\label{eq:stot}
\end{equation}
The first term $(n^{\prime}-n)\sRep$ is the  contribution that is extensive 
in the number of additional replicators. The term $\sRep$ is the entropy
produced during the synthesis of a single replicator (and side products)
from reactants, and its value will depend on the specific physical properties
of the replicator (see Eq.~\eqref{eq:sigmarep} in the Appendix).
The second term, $\ln(n!/n^{\prime}!)$, reflects the entropy increase 
due to the change of the concentration of replicators in solution. 
Eq.~\eqref{eq:stot} holds even when $n^{\prime}<n$,
and even when the right hand side of Eq.~\eqref{eq:stot} is negative
(in which case the transformation $n\to n^{\prime}$ is thermodynamically
disfavored).

To derive Eq.~\eqref{eq:stot} in the Appendix, we introduce several
standard assumptions: (1) replicators and other chemical species
are found in dilute, well-mixed solution, (2) different replicators are statistically indistinguishable from each other, as are the other chemical species, and (3) macrostates corresponding to $1,2,3,\dots$ replicators differ only in terms of the counts
of replicators, reactants, and side products involved in replication.
Importantly, we do not assume that the replicators are in internal
equilibrium, so the result also applies to far-from-equilibrium systems
such as living cells.

We now derive a generalized version of England's bound~\eqref{eq:bnd1}.
For two macrostates containing $n$ and $n+1$ replicators, Eq.~\eqref{eq:stot} gives $\stotBase(n\to n+1)\approx\sRep-\ln(n+1)$.
Assuming first-order growth and decay, the transition probabilities
between these two macrostates may be approximated as $\pi(n\to n+1)\approx ng\,\dtt$
and $\pi(n+1\to n)\approx(n+1)\delta\,\dtt$. The weak form of
LDB~\eqref{eq:ldb2} gives $\stotBase(n\to n+1)\ge\ln\frac{ng}{(n+1)\delta}$.
Combining and simplifying leads to our generalized bound,
\begin{equation}
\sRep\ge\ln\frac{g}{\delta}+\ln n,\label{eq:bnd1b}
\end{equation}
which relates the per-capita dissipation $\sRep$, per-capita growth $g$, per-capita decay $\delta$, and the population size $n$.

Note that in England's analysis, the macrostate $\mII$ (with two replicators) is defined so that the parent and offspring replicators are distinguished, 
and the transition probability $\pi(\mII\to\mI)\approx\delta\,\dtt$
refers only to the decay of the offspring. In our derivation
of Eq.~\eqref{eq:stot}, the replicators are treated as statistically
indistinguishable, and the transition $n+1\to n$ represents the decay
of any of the $n+1$ replicators, occurring with probability $\approx(n+1)\delta\,\dtt$.
The difference is mostly one of convention, not physical content.
For the $1\to 2$ transition ($n=1$), two bounds~\eqref{eq:bnd1} and 
\eqref{eq:bnd1b} are essentially equivalent.

Careful readers may have noticed a thermodynamic inconsistency in
the generalized version of England's bound~\eqref{eq:bnd1b}. This
inconsistency is considered in depth in the next section.

\section{Impossibility theorem}

\label{sec:Impossibility-theorem}

England's bound is usually interpreted as a universal relationship
that relate thermodynamic dissipation to replicator per-capita growth
and decay rates. However, we now argue against the validity of this
interpretation, even in principle. Our critique is based on an ``impossibility
theorem'', which shows that a thermodynamically consistent replicator
cannot undergo both first-order growth and first-order decay back
into reactants.

{There are several equivalent ways to demonstrate our impossibility
result. We do it via two general theoretical arguments, plus
a concrete model of a simple autocatalytic chemical system.

The first (and perhaps simplest) way to demonstrate our impossibility
result is to consider England's bound, in the generalized form~\eqref{eq:bnd1b}
that applies to arbitrary population sizes. A thermodynamically consistent
replicator has a finite per-capita dissipation $\sRep$. However,
the bound~\eqref{eq:bnd1b} holds for all $n\ge0$, which can only
be true if $\sRep=\infty$. In fact, considering the derivation
of this bound, it is only possible to have a finite
$\sRep$ and fixed per-capita growth rate $0<g<\infty$ if
the decay probability scales in a non-first-order manner, as $\pi(n+1\to n)\propto n^{2}$
or faster. Conversely, it is only possible to have a finite
$\sRep$ and fixed per-capita
decay rate $0<\delta<\infty$ if the replication probability does not scale with population
size.}

A second way to demonstrate our impossibility result is to consider
two macrostates: macrostate $\bm{1}$ contains a single replicator
(same as $\mI$ above), while macrostate $\bm{0}$ contains no replicators,
only the reactants needed for replication. Assuming first-order decay,
the transition probability $\pi(\bm{1}\to\bm{0})\approx\delta\,\dtt$
reflects the decay of the single replicator into reactants. The transition
probability $\pi(\bm{0}\to\bm{1})\approx\gamma\,\dtt$ captures the
spontaneous (uncatalyzed) formation of replicator from reactants,
{where we introduced the rate constant $\gamma$ of uncatalyzed formation. Combining Eq.~\eqref{eq:stot}
and the weak form of LDB~\eqref{eq:ldb2} gives a bound on the entropy
produced during the decay process $\bm{1}\to\bm{0}$:}
\begin{equation}
\stotBase(\bm{1}\to\bm{0})\approx-\sRep\ge\ln\frac{\delta}{\gamma}.\label{eq:ldb2-2}
\end{equation}
On the other hand, the generalized England's bound~\eqref{eq:bnd1b}
for $n=1$ implies that $\ln(g/\delta)\le\sRep$. Combining gives
the inequality 
\begin{equation}
\ln\frac{g}{\delta}\le\ln\frac{\gamma}{\delta},\label{eq:ineqss3}
\end{equation}
which implies $g\le\gamma$. However, the defining property
of self-replication is \emph{autocatalysis}, meaning that the formation
of a new replicator in the presence of an existing replicator should
be much faster than spontaneous formation directly from reactants,
\begin{equation}
\gamma\ll g.\label{eq:defrep}
\end{equation}
In fact, perfect first-order growth would require that $\gamma=0$.
More generally, if inequality~\eqref{eq:defrep} did not hold, one
could not interpret macrostate transitions like $n\to n+1$ as replication
events, since the new replicator could instead form spontaneously
from reactants. It is clear that the inequalities~\eqref{eq:ineqss3}
and~\eqref{eq:defrep} are in contradiction with each other.

Finally, 
\global\long\def\xx{n}%
we illustrate our impossibility result using a simple but concrete
model. Consider an autocatalytic chemical reaction such as
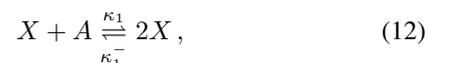
\begin{align}
X+\SS & \underset{\kk_{1}^{-}}{\overset{\kk_{1}}{\rightleftharpoons}}2X\,,\label{eq:autocatalysis-1}
\end{align}
where $X$ is a replicating molecule and $\SS$ is a reactant necessary
for replication. For simplicity, we assume that the reaction is elementary
with mass-action kinetics. We also assume that molecular counts are sufficiently
large such that the system can be described in terms of deterministic
number concentrations, $\xx=[X]$ and $\ss=[\SS]$. Reaction~\eqref{eq:autocatalysis-1}
exhibits forward flux $\kk_{1}\xx\ss$ with forward rate constant
$\kk_{1}$, and reverse flux $\kk_{1}^{-}\xx^{2}$ with backward rate
constant $\kk_{1}^{-}$. We note that the reverse flux is second-order
in $n$. For convenience, we will sometimes use the term \emph{uncopying} 
to refer to the reverse direction of autocatalysis, i.e., the catalyzed
reversion of the replicator back into reactants ($2X\to X+A$).

Suppose that $X$ can also decay back into reactant in an uncatalyzed
fashion,
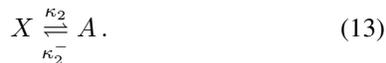
\begin{equation}
X\underset{\kk_{2}^{-}}{\overset{\kk_{2}}{\rightleftharpoons}}\SS\,.\label{eq:decay-1}
\end{equation}
This decay reaction has forward flux $\kk_{2}\xx$ and reverse flux
$\kk_{2}^{-}a$. The two reactions~\eqref{eq:autocatalysis-1} and
\eqref{eq:decay-1} have opposite stoichiometry and therefore opposite
free energy of reaction $-\Delta G$. In the setting of chemical thermodynamics,
local detailed balance implies that $-\Delta G$ (in units of J per
reaction) is equal to the logarithmic ratio of the forward and backward
fluxes \citep{kondepudiModernThermodynamicsHeat2015},
\begin{equation}
-\Delta G=\kB T\ln\frac{\kk_{1}\xx a}{\kk_{1}^{-}\xx^{2}}=\kB T\ln\frac{\kk_{2}^{-}\ss}{\kk_{2}\xx}.\label{eq:stotr}
\end{equation}

In order for the system to exhibit first-order growth rather
than uncatalyzed formation, it must be that $\kk_{1}\xx a\gg\kk_{2}^{-}\ss$,
so that the creation of replicators is dominated by autocatalysis,
not the reverse of the decay reaction. In order for the system to
undergo first-order decay, rather than second-order uncopying, it
must be that $\kk_{2}\xx\gg\kk_{1}^{-}\xx^{2}$. However, these two inequalities are incompatible with Eq.~\eqref{eq:stotr},
highlighting the thermodynamic inconsistency. The underlying reason
is that replication ($X+\SS\to2X$) is thermodynamically favored over
uncopying ($2X\to X+\SS$) to the same extent that uncatalyzed formation
($\SS\to X$) is favored over first-order decay ($X\to\SS$). Thus,
if first-order decay is the dominant pathway for destruction, uncatalyzed
formation must be the dominant pathway for formation.

Of course, if the first-order decay reaction~\eqref{eq:decay-1} occurs
at negligible rates, then the system would exhibit first-order growth
via the forward direction of~\eqref{eq:autocatalysis-1}. In addition,
decay back into reactants would occur due to uncopying, the reverse
direction of the catalyzed reaction~\eqref{eq:autocatalysis-1}. 
However,
in this case, decay will not be first-order (e.g., the elementary
autocatalytic reaction~\eqref{eq:autocatalysis-1} leads to second-order
decay, $\kk_{1}^{-}\xx^{2}$), so it will be inconsistent with the
master equation~\eqref{eq:me}. It will also be inconsistent with
the exponential growth equation~\eqref{eq:exp}, which only holds
for first-order growth and first-order decay.

Above we showed that a thermodynamically consistent replicator cannot
simultaneously exhibit first-order growth and first-order decay back
into reactants. Of course, many replicators do exhibit both first-order
growth and first-order decay. As we discuss in the next section, they
do so by decaying into different waste products, instead of reverting
back into their original reactants.

\section{Alternative degradation pathways}

\label{sec:alt}

Until now, we followed England's analysis in assuming that the decay
transition involves ``reversion of the replicator back into the exact
set of reactants in its environment out of which it was made''. However,
in most replicators of interest, the actual decay process that is
observed is not reversion back into reactants, but rather a separate
degradation process into different waste products. Such replicators
can exhibit both first-order growth and first-order decay. However,
as we argue here, if there is no general relationship between the
processes of replication and decay, then there cannot be a universal
thermodynamic bound that constrains their replication and decay rates.
A related point was raised in an insightful paper by Saakian and Qian
\citep{saakian2016nonlinear}.

As a concrete example, consider again the autocatalytic replicator
\eqref{eq:autocatalysis-1} discussed above. Imagine that the dominant
decay process is neither uncopying, the reverse of reaction~\eqref{eq:autocatalysis-1},
nor uncatalyzed reversion back to reactants, reaction~\eqref{eq:decay-1}.
Rather, decay involves a separate reaction
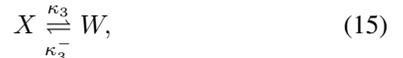
\begin{equation}
X\underset{\kk_{3}^{-}}{\overset{\kk_{3}}{\rightleftharpoons}}W,\label{eq:decay2-1}
\end{equation}
where $W$ is a waste product different from the reactant $\SS$. 

As an example, let us consider the RNA replicator by Lincoln and Joyce
\citep{lincolnSelfsustainedReplicationRNA2009} discussed in England's
paper \citep{england2013statistical}. Here, replication consumes
a reactant RNA molecule with a triphosphate group and releases an
inorganic pyrophosphate as a side product. (The reaction scheme of
the RNA replicator is slightly more complex than elementary scheme
\eqref{eq:autocatalysis-1}, but this does not change the general
point of our argument). Decay can proceed along one of two paths.
The first is the reverse of replication, known as ``pyrophosphorolysis''
\citep{maitraRoleDeoxyribonucleicAcid1967,rozovskaya1982processive,kahn1989reversibility},
in which a pyrophosphate is consumed and a triphosphate-charged RNA
molecule is produced. The second is spontaneous ``hydrolysis'' 
of the RNA phosphodiester bond. Hydrolysis is a separate reaction
that does not involve pyrophosphate and it produces a ``waste''
RNA molecule, with the triphosphate group replaced by a monophosphate
group. 

We use the term \emph{degradation} to refer to the decay of the replicator
into different waste products, as opposed to reversion into the initial
reactants. Because replication and degradation are independent processes,
not reverse directions of the same process, in general they have independent
thermodynamic properties. In such cases, the derivation of Eq.~\eqref{eq:stot}
in the Appendix does not apply, nor does the impossibility result
derived from it in the previous section. Moreover, both replication
and degradation may be thermodynamically favored in the forward direction,
allowing simultaneous first-order growth and first-order degradation.
For instance, for an autocatalytic replicator with reactions~\eqref{eq:autocatalysis-1}
and~\eqref{eq:decay2-1}, the per-capita replication rate may be taken
as $g=\kk_{1}\ss$ (over timescales where the reactant concentration
$\ss$ is approximately constant) and the per-capita degradation rate
may be taken as $\delta=\kk_{3}$.

Notably, when considering actual examples \citep{england2013statistical},
England calculates the decay rate as the rate of degradation into
waste products, rather than the rate of reversion back into reactants.
For example, for the RNA replicator, his estimate is based on the
rate of RNA hydrolysis, not pyrophosphorolysis. For the \emph{E. coli},
his estimate is based on the time required for all peptide bonds in
a single cell to undergo hydrolysis. This differs from the rate of
reversion back into reactants, which would involve the reverse reaction
of protein bond formation, de-respiration of released carbon dioxide
into glucose and oxygen, etc.

Nonetheless, in England's original bound, $\delta$ refers to the rate of reversion back
into original reactants, rather than degradation into other waste
products. To make the connection to degradation, England assumes that
reversion is slower than degradation,
\begin{equation}
\delta\le\delta^{\prime}.\label{eq:lb-2}
\end{equation}
where $\delta^{\prime}$ is the degradation rate. The bound~\eqref{eq:bnd1}
then holds in the weaker form with the reversion rate $\delta$ replaced by the degradation
rate $\delta^{\prime}$, as in $\stotBase\ge\ln({g}/{\delta}^\prime)$. However, there are some problems with this
approach.

For one, the bound may be violated, because there is no \emph{a priori} reason that reversion must be
slower than degradation. For example, for the RNA replicator, England
assumes that hydrolysis (degradation) is faster than pyrophosphorolysis
(reversion), but in fact there is no universal relationship between
the rates of two processes. Moreover, the rate of pyrophosphorolysis
depends on the concentration of pyrophosphate \citep{maitraRoleDeoxyribonucleicAcid1967,rozovskaya1982processive,kahn1989reversibility},
while that of hydrolysis does not. At increased pyrophosphate concentrations,
reversion by pyrophosphorolysis has been observed to proceed as fast as a minute
per nucleotide \citep{rozovskaya1982processive}, orders of magnitude
faster than degradation by hydrolysis (estimated at approximately 4 years per nucleotide \citep{england2013statistical}).

Even for an entire \emph{E. coli} bacterium, it may be debated whether
degradation is always faster than reversion of an offspring cell into
starting reactants.  There are various scenarios that can be imagined
that accelerate reversion; for instance, the parent cell might run
its Krebs cycle in reverse. Of course, there is no doubt that such
reversion is hyper-astronomically unlikely, but one may still wonder
whether it is undeniably \emph{more unlikely} than the hydrolysis of all
peptide bonds, whose probability England estimates at $e^{-6.7\times10^{10}}$
(!) given a 20-minute generation time \citep{england2013statistical}.
In any case, common-sense intuitions about such astronomically unlikely events
should be treated with caution.

The best way to demonstrate that degradation is faster than reversion
is to observe how a replicator actually decays. In many cases, degradation
will be the dominant decay process. However, even in such cases, there
is usually no meaningful thermodynamic constraint on growth and degradation
rates, because the two sides of~\eqref{eq:lb-2} refer to two independent
physical processes and their difference is completely uncontrolled.

Consider again \emph{E. coli} bacteria. 
They are never observed to undergo hydrolysis of all peptide bonds, instead they
simply die at the rate of $\approx5\times10^{-4}$ per generation \citep{wang2010robust}.
This death rate can be related to England's estimate of the entropy
produced during \emph{E. coli} replication, $\approx3.3\times10^{11}$
\citep{england2013statistical}. Plugging these numbers into the bound
\eqref{eq:bnd1} gives
\begin{equation}
\stot=3.3\times10^{11}\ge7.6\approx-\ln(5\times10^{-4}).\label{eq:m2-1-1}
\end{equation}
This inequality is not biologically or physically meaningful because
the two sides differ by a factor of about 50 billion. To put things
in perspective, the inequality tells us that no less than $7.6$ $\kB T$
of free energy must be dissipated in order to replicate a bacterium.
This is a tiny amount, less than the dissipation produced by the hydrolysis
of a single ATP molecule ($\approx20$ $\kB T$).

Above, we argued that the inequality between reversion and degradation
\eqref{eq:lb-2} may be violated, or it may hold but be so weak that
it is irrelevant. Nonetheless, one may wonder whether the reverse
transition probability $\pi(\mII\to\mI)$, as appears in inequality
\eqref{eq:ldb2}, may be defined to account for both reversion and
degradation, so that no further weakening of the bound is necessary.
In fact, whether $\pi(\mII\to\mI)$ accounts for degradation or not
depends in a subtle way on the definition of macrostates $\mI$ and
$\mII$. As an example, consider a replicator that undergoes degradation
into waste species $W$. Now imagine two different ways of defining
these macrostates. Under the first definition, the microstates in
$\mI$ and $\mII$ all contain the same fixed number of waste molecules
$W$. Since degradation increases the number of waste molecules, the
transition $\mII\to\mI$ will not include degradation and the transition
probability $\pi(\mII\to\mI)$ will only account for reversion back
to reactants. Under the second (arguably more realistic) definition,
the precise number of waste molecules varies among different microstates
in $\mI$ and/or $\mII$. Then, the transition probability $\pi(\mII\to\mI)$
will account for both pathways, reversion and degradation.

Nonetheless, under either definition, we end up
with the same very weak thermodynamic bound~\eqref{eq:m2-1-1}. Imagine
that degradation is many orders of magnitude more likely than reversion,
as in the \emph{E. coli} that undergoes degradation by death at the
rate of $\delta^{\prime}=5\times10^{-4}$ per generation. Under the
first definition of the macrostates, $\pi(\mII\to\mI)$ will account
only for reversion and therefore be tiny compared to $\delta^{\prime}\,\dtt$,
so the inequalities~\eqref{eq:lb-2} and~\eqref{eq:m2-1-1} will be
incredibly weak. Under the second definition of the macrostates, $\pi(\mII\to\mI)$
will account for both reversion and decay, and the inequality~\eqref{eq:lb-2}
may be nearly tight. However, the entropy production
$\stot$ and transition probability $\pi(\mI\to\mII)$ that characterize
replication do not depend significantly on whether the degradation waste products are allowed
to fluctuate or not, since these waste molecules are not involved in replication.
Therefore, to the extent that $\pi(\mII\to\mI)$ becomes much larger
and~\eqref{eq:lb-2} tighter, the weak LDB bound~\eqref{eq:ldb2}
will become looser. {This reflects the fact, discussed above
in Section~\ref{subsec:General-result}, that the weak version of
LDB may become very loose when macrostates are internally out of equilibrium
and multiple alternative pathways are present.}

\section{Conclusion}

In this paper, we considered England's bound~\eqref{eq:bnd1} that
relates thermodynamic dissipation $\stot$ to per-capita decay $\delta$ 
and growth $g$ rates. This bound has physical meaning
if decay is the reverse process of replication,
meaning that the offspring replicator reverts back to its original reactants. 
However, as we showed in our impossibility theorem, for a thermodynamically consistent replicator, this reverse
process cannot be first-order, hence $\delta$ cannot be interpreted
as a per-capita decay rate.

Alternatively, the decay rate may be defined as the per-capita rate
of degradation into different waste products $\delta^{\prime}$, rather than
original reactants. In this case, there is no universal 
relationship between the degradation rate $\delta^{\prime}$ and physical properties
of replication, such as the growth rate $g$ and the entropy production
$\stot$. Therefore, the resulting bound~\eqref{eq:bnd1} is not physically
meaningful, and it can even be violated.

\begin{acknowledgments}
I thank Jeremy Owen and Jordi Pi\~nero for useful discussions. This
project has received funding from the European Union\textquoteright s
Horizon 2020 research and innovation programme under the Marie Sk\l odowska-Curie
Grant Agreement No.~101068029.
\end{acknowledgments}

\appendix
\renewcommand{\appendixname}{APPENDIX}
\let\oldsection\section
\renewcommand{\section}[1]{\oldsection{\MakeUppercase{#1}}}

{

\section{Derivation of Eq.~(\ref{eq:stot})}

\global\long\def\FreeNrg{\mathcal{F}}%
\global\long\def\spix{\alpha}%
\global\long\def\pix{i}%
\global\long\def\counts{c}%
\global\long\def\repc{n}%
Here we derive expression~\eqref{eq:stot} for the entropy produced
when transforming a system from a macrostate with $\repc$ replicators
to a macrostate with $\repc^{\prime}$ replicators. We note that similar
results can be found in the literature on chemical thermodynamics
\citep{schmiedlStochasticThermodynamicsChemical2007}. 

We consider a system containing some replicators, which may be chemical
or biological in nature. The system may also contain $K-1$ other
chemical species that serve as additional reactants or side products
of replication. We label the different species using $\spix\in\{1,\dots,K\}$,
with $\spix=1$ indicating the replicator and $\spix\in\{2,\dots,K\}$
indicating reactants and side products.

Each microstate $x$ is written as $x=(\vec{\counts},\vec{y},\vec{s})$,
where $\vec{\counts}=(\counts_{1},\dots,\counts_{K})$ indicates the
vector of population counts of species $\spix\in\{1,\dots,K\}$,
$\vec{y}$ encodes the state of all particles of all species, and
$\vec{s}$ encodes the state of all solvent particles, plus any other
subsystems (catalytic surfaces, etc.) that are not directly involved
in replication as reactant or side product. The vector $\vec{y}=(\vec{y}^{(1)},\dots,\vec{y}^{(K)})$
is further decomposed into vectors $\vec{y}^{(\spix)}$ representing
particles of each species, and each $\vec{y}^{(\spix)}$
is further decomposed as $\vec{y}^{(\spix)}=(\vec{y}^{(\spix,1)},\dots,\vec{y}^{(\spix,\counts_{\alpha})})$,
where $\vec{y}^{(\spix,\pix)}$ encodes the state of particle $\pix$
of species $\alpha$. Specifically, $\vec{y}^{(\spix,\pix)}$ encodes
the position of the particle within the reactor volume, plus internal
degrees of freedom (e.g., configurational, rotational, and vibrational
modes, internal structure in the case of complex particles, etc.).
The order of the indices $\pix$ of the particles of each species
is arbitrary, so any $\vec{y}^{(\spix)}$ and $\vec{y}^{(\spix)\prime}$
that differ only in the ordering of particles are equivalent.

The macrostate containing $\repc$ replicators is represented by $P_{\repc}(x)\equiv P_{\repc}(\vec{\counts},\vec{y},\vec{s})$,
a distribution over microstates with support restricted to microstates
with $\counts_{1}=\repc$. 

We introduce several standard assumptions. First, the macrostates
$P_{\repc}$ differ only in the number of replicators and reactant/product
species, while other statistical properties are the same. Specifically,
the statistical properties of the solvent are the same, $P_{\repc}(\vec{s})\equiv P(\vec{s})$,
and so are the conditional probabilities of population counts, up
to differences due to consumption/creation of species by replicator
synthesis:
\begin{equation}
P_{\repc}(\vec{\counts}\vert\vec{s})=P_{0}(\vec{\counts}-\repc\vec{\Delta}\vert\vec{s}).\label{eq:pmUpdated}
\end{equation}
The vector $\vec{\Delta}$ specifies the stoichiometry of replicator synthesis,
where $\Delta_{\alpha}>0$ is the number of $\spix$
created as side products and $\Delta_{\alpha}<0$ is the number of
$\spix$ consumed as reactant. By definition, $\Delta_{1}=1$ (the
synthesis of a replicator creates one replicator).

In addition, we assume that the particles of species $\spix\in\{1,\dots,K\}$
are in dilute, well-mixed solution without long-range interactions.
Then, the energy of a typical microstate $x=(\vec{\counts},\vec{y},\vec{s})$
is approximately additive,
\begin{equation}
E(\vec{\counts},\vec{y},\vec{s})\approx E_{\varnothing}(\vec{s})+\sum_{\alpha=1}^{K}\sum_{\pix=1}^{\counts_{\spix}}E_{\alpha}(\vec{y}^{(\spix,\pix)},\vec{s}).\label{eq:nrg0}
\end{equation}
Here $E_{\varnothing}(\vec{s})$ is the solvent energy and $E_{\alpha}(\,\cdot\,,\vec{s})$
is the single-particle energy for species $\spix$, which may account
for particle-solvent interactions. Assumptions of well-mixedness imply that
correlations between particles are negligible, once conditioned on
the state of the solvent $\vec{s}$. Thus, the probability of any microstate
$x=(\vec{\counts},\vec{y},\vec{s})$ factors like
\begin{align}
P_{\repc}(\vec{\counts},\vec{y},\vec{s}) & \approx P(\vec{s})P_{\repc}(\vec{\counts}\vert\vec{s})\prod_{\spix=1}^{K}\prod_{\pix=1}^{\counts_{\spix}}P_{\repc,\alpha,\pix}(\vec{y}^{(\spix,\pix)}\vert\vec{s})\label{eq:dist0}
\end{align}
where $P_{\repc,\alpha,\pix}$ is the conditional probability distribution
of particle $\pix$ of species $\alpha$ in a system
with $\repc$ replicators.

Finally, we assume that particles in each species have indistinguishable
statistical properties,
\begin{equation}
P_{\repc,\alpha,\pix}(\,\cdot\, \vert\vec{s},\vec{\counts})=\omega_{\spix}(\,\cdot \,\vert\vec{s}).\label{eq:dist1}
\end{equation}
The conditional distribution $\omega_{\spix}$ encodes the statistical
properties of a single particle of species $\spix$, and it does not
depend on particle label $\pix$, replicator count $\repc$, nor the counts of other species. Importantly,
we do not assume that the distribution $\omega_{\alpha}$ has a Boltzmann
form. Thus, the internal state of the replicators and other
species may be arbitrarily far-from-equilibrium. For instance,
when studying biological replicators, $\omega_{1}$ might represent
the highly nonequilibrium steady-state distribution of a living bacterium.

We now evaluate the entropy produced when the system is transformed
from macrostate $P_{\repc}$ to macrostate $P_{\repc^{\prime}}$.
As in Eq.~\eqref{eq:stot2}, the entropy production is given by 
\begin{equation}
\stotBase(\repc\to\repc^{\prime})=[S(P_{n^{\prime}})-S(P_{n})]+\langle Q\rangle_{n\to n^{\prime}}/k_{B}T,\label{eq:stotapp}
\end{equation}
where the heat is equal to the change of loss of expected energy,
$\langle Q\rangle_{n\to n^{\prime}}=\langle E\rangle_{P_{\repc}}-\langle E\rangle_{P_{\repc^{\prime}}}$.

To calculate the expected energy and entropy, it will be useful to consider
the energy and entropy contribution from a single particle of species
$\alpha$ as
\begin{align*}
\mathcal{E}_{\alpha} & :=\int P(\vec{s})\omega_{\spix}(\vec{z}\vert\vec{s})E_{\alpha}(\vec{z},\vec{s})\,d\vec{z}\\
\mathcal{S}_{\alpha} & :=-\int P(\vec{s})\omega_{\spix}(\vec{z}\vert\vec{s})\ln\omega_{\spix}(\vec{z}\vert\vec{s})\,d\vec{z}.
\end{align*}
Using the above definitions and assumptions, we may write the expected
energy under $P_{n}$ as
\begin{align}
 & \langle E\rangle_{P_{\repc}}\approx\langle E_{\varnothing}(\vec{s})\rangle_{P}+\int d\vec{s}\sum_{\alpha=1}^K\sum_{\counts_{\alpha}=1}P_{n}(\counts_{\alpha},\vec{s})\counts_{\alpha}\langle E_{\alpha}\rangle_{\omega_{\spix}(\cdot\vert\vec{s})}\nonumber \\
 & =\langle E_{\varnothing}(\vec{s})\rangle_{P}+\int d\vec{s}\sum_{\alpha=1}^K\sum_{\counts_{\alpha}=1}P_{0}(\counts_{\alpha},\vec{s})(\counts_{\alpha}+\repc\Delta_{\alpha})\langle E_{\alpha}\rangle_{\omega_{\spix}(\cdot\vert\vec{s})}\nonumber \\
 & =\langle E\rangle_{P_{0}}+\repc\sum_{\alpha=1}^K\Delta_{\alpha}\mathcal{E}_{\alpha}\nonumber \\
& =\langle E\rangle_{P_{0}}+n\Big[\mathcal{E}_{1}+\sum_{\alpha=2}^K\Delta_{\alpha}\mathcal{E}_{\alpha}\Big].
\label{eq:Eexp}
\end{align}
In the second line, we performed the change of variables $\counts_{\alpha}\mapsto\counts_{\alpha}+\repc\Delta_{\alpha}$,
then used Eq.~\eqref{eq:pmUpdated}. 

To compute the Shannon entropy,
we write
\begin{align}
 & S(P_{n})=S[P(\vec{s})]+S[P_{\repc}(\vec{\counts}\vert\vec{s})]+S[P_{\repc}(\vec{y}\vert\vec{\counts},\vec{s})]\nonumber \\
 & \quad\approx S[P(\vec{s})]+S[P_{0}(\vec{\counts}\vert\vec{s})]+\sum_{\alpha=1}^K S[P_{\repc}(\vec{y}^{(\alpha)}\vert\vec{\counts},\vec{s})],\label{eq:f1}
\end{align}
where we first used the chain rule for Shannon entropy, $S[P_{n}(\vec{\counts}\vert\vec{s})]=S[P_{0}(\vec{\counts}-\repc\vec{\Delta}\vert\vec{s})]=S[P_{0}(\vec{\counts}\vert\vec{s})]$ and Eq.~\eqref{eq:dist0}.
For each species $\alpha$, we have
\begin{align*}
 & S[P_{\repc}(\vec{y}^{(\alpha)}\vert\vec{\counts},\vec{s})]=
 \\
 & \int d\vec{s}\,\sum_{\counts_{\alpha}=1}P_{\repc}(\counts_{\alpha},\vec{s})\big[\counts_{\alpha}\langle-\ln\omega_{\spix}(\vec{z}\vert\vec{s})\rangle_{\omega_{\spix}(\cdot\vert\vec{s})}-\ln\counts_{\alpha}!\big],
\end{align*}
where $-\ln\counts_{\spix}!$ accounts for the fact that the indexing
order of the particles does not matter. We further rewrite the right side,
\begin{align*}
 & \int d\vec{s}\,\sum_{\counts_{\alpha}}P_{0}(\counts_{\alpha},\vec{s})\times\nonumber\\
 & \quad \big[(\counts_{\alpha}+\repc\Delta_{\alpha})\langle-\ln\omega_{\spix}(\vec{z}\vert\vec{s})\rangle_{\omega_{\spix}(\cdot\vert\vec{s})}-\ln(\counts_{\alpha}+\repc\Delta_{\alpha})!\big]\nonumber\\
 &=S[P_{0}(\vec{y}^{(\alpha)}\vert\vec{\counts},\vec{s})]+ n\Delta_{\alpha}\mathcal{S}_{\alpha}+\Big\langle\ln\frac{c_{\alpha}!}{(\counts_{\alpha}+\repc\Delta_{\alpha})!}\Big\rangle_{P_{0}}
\end{align*}
where we again performed the change of variables $\counts_{\alpha}\mapsto\counts_{\alpha}+\repc\Delta_{\alpha}$.
Combining  with Eq.~\eqref{eq:f1} gives
\begin{align}
 & S(P_{n})= S(P_{0})+n\Delta_{\alpha}\mathcal{S}_{\alpha}+\Big\langle\ln\frac{c_{\alpha}!}{(\counts_{\alpha}+\repc\Delta_{\alpha})!}\Big\rangle_{P_{0}}.\label{eq:f2}
\end{align}

We simplify the factorial terms in Eq.~\eqref{eq:f2} in the following manner.  
For the replicator
species $\alpha=1$, we write
\[
\left\langle \ln\frac{\counts_{1}!}{(\counts_{1}+\repc\Delta_{1})!}\right\rangle _{P_{0}}=-\ln\repc!
\]
which follows from $\Delta_{1}=1$ and $\counts_{1}=0$ for all microstates
with support under $P_{0}$. For non-replicator species $\alpha>1$,
we assume that their counts are very large, relative to the number consumed or produced
by replicator synthesis. We then apply Stirling's approximation $\ln x!\approx x\ln x-x$
and simplify,
\begin{align*}
 & \ln\frac{\counts_{\alpha}!}{(\counts_{\spix}+\repc\Delta_{\spix})!}\\
 & \quad\approx\repc\Delta_{\alpha}-\counts_{\alpha}\ln\frac{\counts_{\alpha}+\repc\Delta_{\alpha}}{\counts_{\alpha}}-\repc\Delta_{\alpha}\ln(\counts_{\alpha}+\repc\Delta_{\alpha})\\
 & \quad\approx-\repc\Delta_{\alpha}\ln(\counts_{\alpha}+\repc\Delta_{\alpha})\\
 & \quad\approx-\repc\Delta_{\alpha}\ln\counts_{\alpha}.
\end{align*}
These approximations are valid when $\counts_{\spix}\gg\vert\repc\Delta_{\spix}\vert$. 

Finally, combining with Eq.~\eqref{eq:f2} gives
\[
S(P_{\repc})\approx S(P_{0})+\repc\left[\mathcal{S}_{1}+\sum_{\alpha=2}^K\Delta_{\alpha}(\mathcal{S}_{\alpha}-\langle\ln\counts_{\alpha}\rangle_{P_{0}})\right]-\ln\repc!
\]
Plugging this and Eq.~\eqref{eq:Eexp} into Eq.~\eqref{eq:stotapp} gives the expression~\eqref{eq:stot} in the main text,
with per-capita entropy production
\begin{equation}
\sRep\approx \mathcal{S}_{1}-\frac{\mathcal{E}_{1}}{k_{B}T}+\sum_{\alpha=2}^K\Delta_{\alpha}\Big(\mathcal{S}_{\alpha}-\langle\ln\counts_{\alpha}\rangle_{P_{0}}-\frac{\mathcal{E}_{\alpha}}{k_{B}T}\Big).\label{eq:sigmarep}
\end{equation}
}

\vfill

\bibliographystyle{IEEEtran}
\bibliography{writeup}

\clearpage
\end{document}